\documentclass[journal,10pt,twocolumn]{IEEEtran}
\usepackage[table]{xcolor}
\usepackage{multicol}
\usepackage{color, colortbl}
\usepackage[first=0,last=9]{lcg}

\usepackage{cite}
\usepackage[english]{babel}
\usepackage{multicol}
\usepackage{multirow} 
\usepackage{amsmath}
\usepackage{amsmath}
\usepackage{amsfonts}
\usepackage{amssymb}
\usepackage{float} 
\usepackage{amsmath}
\usepackage{color}
\usepackage{array}
\usepackage{graphicx}
\usepackage{epstopdf}
\usepackage{epsfig}
\usepackage{algorithm}
\usepackage{textcomp}
\usepackage[noend]{algorithmic}
\usepackage{url}
\usepackage{balance}
\usepackage{hhline}
\usepackage{comment}
\usepackage[numbers, square, comma, sort&compress]{natbib}

\usepackage[singlespacing]{setspace}
\usepackage[font=footnotesize,labelfont=rm,figurename=Fig.]{caption}
\usepackage{etoolbox}
\usepackage{mathtools}
\usepackage[many]{tcolorbox}
\usepackage{lipsum}

\makeatletter
\let\@nodottedtocline\@dottedtocline
\patchcmd{\@nodottedtocline}{\hbox{.}}{\hbox{}}{}{}
\patchcmd{\@nodottedtocline}{\normalcolor #5}{\normalcolor}{}{}
\newcommand*\l@sectionsubtitle{\@nodottedtocline{1}{0em}{1.5em}}
\makeatother

\newcounter{inlineequation}
\DeclareMathAlphabet{\mathpzc}{OT1}{pzc}{m}{it}
\definecolor{Gray}{gray}{0.9}
\definecolor{LightCyan}{rgb}{0.88,1,1}

\NewTColorBox{NewBox}{ s O{!htbp} }{%
  floatplacement={#2},
  IfBooleanTF={#1}{float*,width=\textwidth}{float},
  colframe=white,colback=blue!10!white,
  }

\begin{document}
\title{FSO-based Vertical Backhaul/Fronthaul Framework for 5G+ Wireless Networks}

\author{\IEEEauthorblockN{Mohamed Alzenad, Muhammad Z. Shakir, Halim  Yanikomeroglu, and Mohamed-Slim Alouini
\thanks{M. Alzenad and H. Yanikomeroglu are with the Department of Systems and Computer Engineering, Carleton University, Ottawa, Canada, Email: \{mohamed.alzenad,halim\}@sce.carleton.ca.}
\thanks{M. Z. Shakir is with the School of Engineering and Computing, University of the West of Scotland, Paisley, Scotland, UK, Email: muhammad.shakir@uws.ac.uk.}
\thanks{M.-S. Alouini is with the Computer, Electrical and Mathematical Sciences and Engineering Division, King Abdullah University of Science and Technology, Thuwal, Saudi Arabia, Email: slim.alouini@kaust.edu.sa.}
}


}

\maketitle
\thispagestyle{empty}

\begin{abstract}

The presence of a super high rate, but also cost-efficient, easy-to-deploy, and scalable, backhaul/fronthaul framework is essential in the upcoming fifth-generation (5G) wireless networks \& beyond. Motivated by the mounting interest in the unmanned flying platforms of various types including unmanned aerial vehicles (UAVs), drones, balloons, and high-altitude/medium-altitude/low-altitude platforms (HAPs/MAPs/LAPs), which we refer to as the networked flying platforms (NFPs), for providing communications services and the recent advances in free-space optics (FSO), this article investigates the feasibility of a novel vertical backhaul/fronthaul framework where the NFPs transport the backhaul/fronthaul traffic between the access and core networks via point-to-point FSO links. The performance of the proposed innovative approach is investigated under different weather conditions and a broad range of system parameters. Simulation results demonstrate that the FSO-based vertical backhaul/fronthaul framework can offer data rates higher than the baseline alternatives, and thus can be considered as a promising solution to the emerging backhaul/fronthaul requirements of the 5G+ wireless networks, particularly in the presence of ultra-dense heterogeneous small cells. The paper also presents the challenges that accompany such a novel framework and provides some key ideas towards overcoming these challenges.
\end{abstract}


\section{Introduction}

It is widely acknowledged that one of the key architectural enablers  towards the extremely high data rate coverage in wireless networks is the dense deployment of small cells. Although the small cell concept has been envisioned and studied for many years within the 4G LTE  framework, the concept has never found widespread application mainly  due to the cost of deployment. In the conventional wireless networks, the cost of the macro cell base station (macro-BS) has been a dominant factor. The cost of a small cell base station (SBS), on the other hand, is much lower in comparison to  that of a macro-BS; but the difficulty and cost in  backhauling/fronthauling a high number of SBSs has emerged as a  significant challenge. At the time of writing this article, the 5G standardization process has already started. The 5G networks are  expected to be deployed starting in approximately 2020; the small cell  concept is still perceived as a key 5G enabler. However, the efficient  backhauling/fronthauling of the SBSs remains to be a significant challenge.

This paper aims at exploring a novel radio access network (RAN) architecture to realize a dense small cell deployment, in which SBSs  are connected to the core network through a vertical backhaul/fronthaul; the key technologies within this novel framework are the free-space optics (FSO) and networked flying platforms (NFPs). In this article, we adopt the generic term NFPs to encompass the floating and moving aspects of the unmanned flying platforms of various types including unmanned aerial vehicles (UAVs), drones, balloons, and high-altitude/medium-altitude/low-altitude platforms (HAPs/MAPs/LAPs). 
It is worth noting that although the explored RAN architecture may have a higher cost in comparison to other competing existing  solutions according to the pricing estimate at the time of writing of this  article, however, the sharp reduction expected in the cost of FSO as well as in the operation of NFPs may make the studied novel approach a viable  solution in the next 10 years. At that time, 5G networks will likely  be fully operational, and the standardization of the evolved 5G and  even perhaps 6G networks will have started. For these reasons, in this  paper we are using the generic term 5G+ to denote the 5G and beyond-5G  wireless networks of the future.
\begin{figure*}[t]
\begin{center}
\includegraphics[height=10cm, width=18cm]{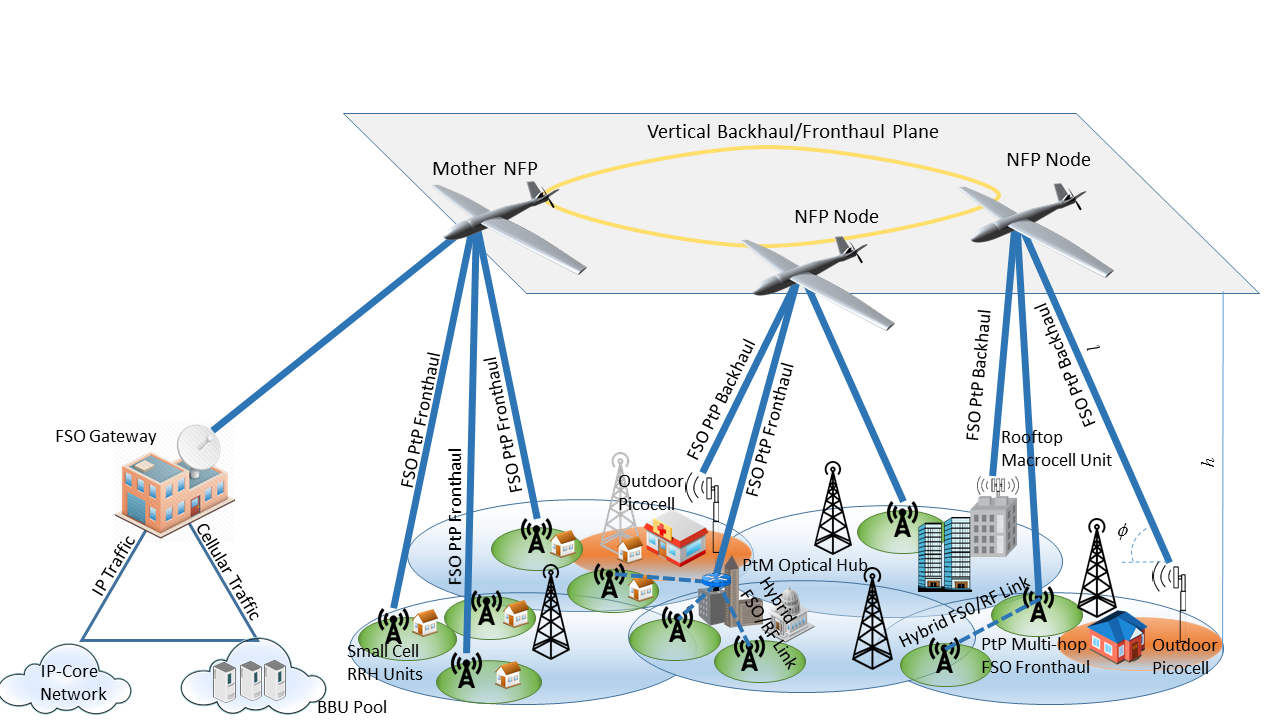}
\caption{\footnotesize Graphical illustration of vertical fronthaul/backhaul framework for 5G+ wireless networks.}
\label{fig1}
\end{center}
\end{figure*} 

\subsection{\textbf{Overview of Backhauling/Fronthauling}}
Two fundamentally different backhauling/fronthauling possibilities exist, namely, wired and wireless backhaul/fronthaul networks. Wired backhaul/fronthaul solutions include copper and fiber. One may confidently state that fiber is always the best option. However, fiber does not always exist at the potential locations of the SBSs, and installing new fiber for small cells may not be an acceptable solution due to the high cost in many environments \cite{Dahrouj}.

A more cost-effective and easy-to-deploy alternative is wireless backhaul/fronthaul, where small cells traffic is carried over microwave links or FSO links. The microwave backhaul relies on particular frequency bands in the range 6-60 GHz. However, these frequency bands are becoming congested in a number of countries \cite{Siddique}. Moreover, line-of-sight (LOS) FSO has recently gained attention as it relies on license-free point-to-point (PtP) narrow beams. 

Current backhauling/fronthauling solutions are based on delivering the traffic of small cells to an aggregation point (central hub). The optimal hub placement problem has been shown to be NP-complete; as a matter of fact, the LoS hub placement may turn out to be simply infeasible since with ultra-dense small cell deployment, particularly in ultra-dense urban areas, some small cells are deployed in hard to reach areas in which LOS propagation is impossible \cite{Karamad}. In such scenarios, RF non-LOS point-to-multipoint (NLOS PtM) solutions which rely on licensed sub-6 GHz spectrum or microwave frequencies could be used at the expense of severe interference, congestion and higher cost \cite{Siddique}. Therefore, a novel paradigm shift of backhaul/fronthaul network design for 5G networks and beyond is needed.

NFPs have recently gained great attention in the communications sector \cite{Marcus}. They can, for instance, deliver cellular and Internet services to remote and dedicated regions where infrastructure is not available and expensive to deploy. 
Recently, Facebook's project Internet.org has been launched to provide Internet coverage via stratosphere communications\footnote {Visit \url{http://info.internet.org } for further information about the project.}. Another example is the recently envisioned drone-BS concept \cite{bor2016new}.

This article particularly addresses the following questions:
\begin{itemize}

\item Can 5G+ networks have a vertical backhaul/fronthaul network based on NFPs and FSO?

\item What are the challenges in integrating FSO-based NFPs into the backhaul/fronthaul network?
\end{itemize}

\begin{table*}[t!]
\caption{Fog attenuation for different wavelengths and foggy conditions.}
\centering
\begin{tabular}{|c|c|c|c|c|c|}
\hline
\multirow{2}{*}{} &\multicolumn{5}{|c|}{Foggy conditions}\\
 \hhline{~-----}
& Dense & Thick & Moderate & Light & Very light\\
\hline
\multirow{1}{*}{Visibility (m)} & 50 & 200	& 500	& 770	 & 1900\\
\hline
\multirow{1}{*}{Wavelength (nm)} &\multicolumn{5}{|c|}{Attenuation dB/km}\\
\hline
650 & 327.61	& 80.19	& 31.43	& 20.16	& 7.92\\
\hline
850 & 309.21	& 73.16	& 27.75	& 17.46	& 6.52\\
\hline
1330 & 280.77	& 62.77	& 22.54	& 13.73	& 4.71\\
\hline
1550 & 271.66	& 59.57	& 20.99	& 12.65	& 4.22\\
\hline
\end{tabular}
\end{table*}

\subsection{\textbf{Overview of Networked-Flying Platforms}}

Different terms are used for unmanned aircrafts in literature and practice. The term unmanned aerial vehicle (UAV) often refers to the flying platform including its payload while the term unmanned aerial system (UAS) refers to the flying platform and the ground station that controls the aerial platform \cite{austin}. The terms low altitude platform (LAP), medium altitude platform (MAP), and high altitude platform (HAP) are mainly used for quasi-stationary objects such as unmanned airships or balloons. 
In this article, we consider NFPs that can carry heavy payloads (FSO transceivers),
float in the air at a quasi-stationary position with the ability to move horizontally and vertically, and offer wireless links to backhaul/fronthaul SBSs. The reason for such moving requirement is that the NFPs may need to float most of the time and may only move based on weather conditions, coverage requirements, and even due to some real-time traffic changes/abnormalities in the network.
 
A NFP can either fly autonomously or non-autonomously. In non-autonomous operation, the NFP is controlled remotely by a human operator on the ground. In autonomous operation, the NFP can perform the task without the need for direct human control. The NFPs can either operate in a single NFP mode or a swarm of NFPs. In a single NFP operation, the mission is performed by a single NFP with no cooperation with other NFPs, if they exist. NFP swarm operation allows multiple interconnected NFPs to cooperate and perform the mission autonomously with one of the NFPs acting as the lead.

\subsection{\textbf{Our Contributions}}
The main contributions of this article are summarized as follows:
\begin{itemize}
\item We investigate the feasibility of a novel vertical backhaul/fronthaul framework that integrates NFPs with FSO technology.
\item We investigate the impact of several weather conditions on the data rates offered by a vertical FSO link.
\item We illustrate the deployment and operational cost of the vertical FSO-based system compared to terrestrial backhauling/fronthauling solutions.
\item We present an overview of solutions for future implementation to increase achievable data rate and link margin under bad weather conditions.
\end{itemize}

\section{Proposed System}

Consider a multi-tier HetNet where ultra-dense small cells overlaid with macro cells within some geographical area. We assume that our system is complementary to the terrestrial backhaul/fronthaul network, and needs to be deployed in many challenging environments, e.g., due to a failure in the terrestrial transport network or a temporarily demand for backhaul/fronthaul during a social event such as a sports event. Moreover, our system is also capable of offering backhaul/fronthaul to the SBSs that are located in hard to reach areas where fiber or microwave links may not readily available and expensive to deploy. Examples of these locations could be in rural/remote areas or urban areas below surrounding buildings where SBSs are expected to be deployed closer to street levels and mounted on street furniture such as walls and lamb posts. 

The proposed backhaul/fronthaul network utilizes NFPs and FSO technology as a transport network that links the SBSs and the core network. Fig. \ref{fig1} shows a graphical illustration of the proposed system referred to as vertical backhaul/fronthaul framework for 5G+ wireless networks. The proposed system has a vertical backhaul/fronthaul plane that comprises of several connected NFPs, we refer to them as the NFP nodes. The flying altitude $h$ could range from a few hundred meters to several kilometers (typically 20 km), depending on coverage area, weather conditions and the NFP's serving capability\footnote{The probability of LoS ($\textup P_{\textup{LoS}}$) connection between a ground point and a NFP increases as the NFP altitude increases. Therefore, the NFP should fly at an altitude high enough to guarantee a LoS connection.}. The NFPs are assumed to fly autonomously in a swarm of NFPs mode. The swarm is controlled by a NFP called a mother NFP which also connects the swarm of NFPs to the core network on the ground. The NFP to mother NFP and mother NFP to core network connectivity are based on FSO links. The connectivity between SBSs or aggregation points and the NFPs is also based on PtP FSO beams. Each SBS with a clear LOS link with a NFP delivers/receives its uplink/downlink traffic to/from the intended NFP via FSO link\footnote{Such FSO transceiver for flying platforms is already available, e.g., MLT-20 from Vialight, which is ideal for flying platforms. Visit \url{http://www.vialight.de/} for further information about the product.}. On the other hand, SBSs with no LOS with a NFP can be served by a nearby SBS which has a clear LOS with a NFP. 
As a further option, instead of connecting every single SBS to a NFP, traffic is backhauled/fronthauled from SBSs to a NFP in a distributed manner. The traffic of the SBSs is relayed to a specified SBS where an aggregation hub is located, and a LOS with a NFP is available. This distributed solution has been shown to have a higher energy efficiency than a centralized solution where the SBSs traffic is delivered to a macro-BS \cite{Ge}. Similarly, SBSs served by NFPs are equipped with tracking systems to automatically point the beam to the desired NFP. NFPs are equipped with steerable FSO units with optical beam steering devices to establish wireless links with the radio access technologies or hub points on ground via FSO beams. 


\section{Link Budget Analysis of Vertical Backhaul/Fronthaul Framework}

FSO is an optical technology that relies on laser to send a very high bandwidth information data. The optical signal is vulnerable to atmospheric attenuation (absorption, scattering and turbulence) and other losses such as geometrical, pointing and optical losses.

\subsection{\textbf{Absorption Loss}}

Absorption occurs when the photons of the FSO beams collide with gaseous molecules which convert the photons into kinetic energy. Absorption depends on the type of gas molecules and their concentration as well as on the transmission frequency of the optical beam.  The wavelength selectivity of absorption allows specific frequencies to pass through it which results in transmission windows in which the absorption loss is negligible \cite{Kaushal}.   

\subsection{\textbf{Scattering Loss}}

Scattering occurs when the FSO beam collides with the particles in the atmosphere. The atmosphere is the layer of gases that surround the planet Earth, and divided into layers according to major changes in temperature. The bottom two layers above the earth surface where the NFPs could fly are the Troposphere (0 to 12 km) and the Stratosphere (12 to 50 km). However, in this article, we consider NFPs that fly at altitude from several hundred meters to several kilometers (less than 20 km). Scattering can be classified into three categories, namely, (1) Rayleigh scattering, (2) Mie scattering, (3) Non-selective scattering. Rayleigh scattering occurs when the wavelength of the FSO beam is large compared to the size of the particles such as air molecules. Mie scattering occurs for particles that are of similar size to the wavelength of the FSO beam such as aerosol particles, fog and haze.  Non-selective scattering occurs when the radius of the particles, such as raindrops, is much larger than the wavelength of the FSO beam. Among the three scattering mechanisms, the FSO beam is mostly attenuated by Mie scattering. 
 The Kruse model describes the attenuation due to Mie scattering as \cite{Grabner}:

\begin{equation}
L_{sca}=4.34\;\beta_{sca}d=4.34\;\left(\frac{3.91}{V}{\Big({\frac{\lambda}{550}}}\Big)^{-\delta}\right)d,
\end{equation}
where $L_{sca}$ denotes the attenuation in dB, $\beta_{sca}$ stands for the scattering coefficient in $\text{km}^{-1}$ and $d$ represents the distance along which the scattering phenomena occurs in km.
 Variable $V$ denotes the visibility range in km, $\lambda$ stands for the transmission wavelength in nm, $\delta= 0.585\;V^{(1/3)}$ for $V<6$ km, $\delta=1.3$ for $6<V<50$ km, and $\delta=1.6$ for $V>50$ km.

\subsubsection{\textbf{Fog Attenuation}}

The FSO beam is highly affected by fog as it causes Mie scattering. Equation (1) can be used to predict the fog attenuation ($L_{fog}$) and $d=\Delta d_{fog}/\textup{sin}(\phi)$, where $\Delta d_{fog}$ is the fog layer thickness. As an example, Table I shows the relationship between different foggy conditions, visibility, wavelength and attenuation per km for selected operating wavelengths.
\subsubsection{\textbf{Rain Attenuation}}

The rainfall causes a non-selective scattering. The rain attenuation, $L_{rain}$ (measured in dB) is given by $L_{rain}=1.076\;R_{rain}^{0.67}d_{rain}$, where $R_{rain}$ denotes the rainfall rate in mm/hour and $d_{rain}$ denotes the distance along which the rain affects the FSO beam in km, and given by $d_{rain}=\Delta d_{rain}/\textup{sin}(\phi)$, where $\Delta d_{rain}$ is the rain layer thickness and $\phi$ is the elevation angle \cite{Muhammad}. 

\subsubsection{\textbf{Cloud Attenuation}}

Clouds are a large collection of a very small and light water droplets or ice crystals. Clouds can be characterized by their height, number density $(N_{d})$, liquid water contents (LWC), water droplet size and horizontal distribution extent. Water droplets in clouds vary in density from 100 to 500 water droplets per cubic centimeters and their water content vary from $3.128\times 10^{-4}$  $\text{g/m}^3$ for a thin Cirrus cloud to 1 $\text{g/m}^3$ for  a Cumulus cloud \cite{Awan}. Different empirical approaches have been proposed to model the cloud attenuation ($L_{cloud}$). In this article, we adopt the approach developed in \cite{Awan}. The approach is based on estimating cloud visibility range by dividing the atmosphere into layers. Then, for each layer, visibility range is estimated from their $N_{d}$ and LWC, where visibility range is given by $V=1.002 (\text{LWC}) N_{d}^{-0.6473}$. Then, (1) is used to predict the cloud attenuation.

\subsection{\textbf{Turbulence Loss}}

The refractive index structure parameter $C_n^2(h)$ is an altitude dependent measure of the turbulence strength. Based on measurements, various models are available to predict the parameter $C_n^2(h)$ such as Hufnagel-Valley (H-V) model \cite{Epple}. According to the H-V model, the parameter $C_n^2(h)$ for the vertical link in the proposed system is given by

\begin{multline}
C_n^2(h)=0.00594\left(\frac{v}{27}\right)^2\left(10^{-5}h\right)^{10}\text{exp}\left(\frac{-h}{1000}\right)\\+2.7\times 10^{-16}\text{exp}\left(\frac{-h}{1500}\right)+A\: \text{exp}\left(\frac{-h}{100}\right),
\end{multline}
where $v$ denotes the rms wind speed and a typical value for constant $A$ is  $1.7\times 10^{-14}  \text{m}^{-2/3}$. The attenuation caused by scintillation, $L_{sci}$ (in dB) is then given by $L_{sci}=2\; \sqrt[]{23.17(\frac{2\pi}{\lambda}10^9)^{\frac{7}{6}}C_n^2(h)\:l^{\frac{11}{6}}}$, where $l$ is the path length \cite{Muhammad}.

\begin{NewBox}[t]
\textbf{Geometrical loss:} 
The geometrical loss in dB is given by $L_{geo}=10\text{log}(\frac {A_R}{A_B})$, where $A_R$ and $A_B$ denote the area of the FSO receiver and beam, respectively. Assuming the receiver has a circular mirror with a radius $r$, then, $A_R=\pi r^2$. The area of the beam at the receiver is a circular disc with some diameter $d_B$ that depends on the length of the communication link $l$ and the divergence angle of the transmitter $\theta$, and it is given by $d_B=\theta l$. The radius of the beam at the receiver is given by $r_B=\frac{d_B}{2}=\theta l/2$. The area of the beam at the receiver is then $A_B=\pi r_B^2$. The geometrical loss is then given by $L_{\textup {geo}}=10\text{log}\left(\frac {\pi r^2}{\pi(\theta l/2)^2}\right)$.
\end{NewBox}

\subsection{\textbf{Geometrical Loss}}
As the light travels through the atmosphere, the light energy spreads out over a larger area. This in turn reduces the power collected by the receiver. The geometrical loss in dB is given by $L_{\textup {geo}}=10\text{log}\left(\frac {\pi r^2}{\pi(\theta l/2)^2}\right)$, where $r$ is the radius of the receiver's aperture, $l$ is the length of the communication link, and $\theta$ is the divergence angle of the transmitter. 

\subsection{\textbf{Pointing and Optical Losses}}

The FSO link is a point-to-point link. Moving NFPs at high altitudes and under extreme turbulence conditions may cause higher pointing losses $(L_{poi})$ that could result in a link failure or a significant reduction in the received signal power\footnote{Recently, some emerging FSO systems are designed with tracking systems using electro-optic or acousto-optic devices that especially applicable to the fast moving platforms for compensating the pointing losses \cite{stamatios}.}. On the other hand, optical losses $(L_{opt})$ is caused due to the imperfect optical elements used at the FSO transceiver which reduces the optical efficiency of the FSO transmitter $(\eta_t)$ and the receiver $(\eta_r)$. The optical losses in dB is given by $L_{opt}=10\text{log}(\eta_t\eta_r)$.



\section{Performance Evaluation of the Proposed Vertical Link}
In this section, first, we present the experimental results of the system under different weather conditions. Second, we discuss some economics associated with the underlying system.

\subsection{\textbf{Data Rate and Link Margin}} 

The vertical FSO links can support a single SBS or multiple SBSs. In the case of supporting multiple SBSs, traffic is backhauled/fronthauled from multiple SBSs to an aggregation point via wireless/wired high capacity links. The number of aggregated SBSs that a single FSO link can support depends on the achievable data rate on the FSO link $R$ and the estimated backhaul/fronthaul traffic for the small cells. The backhaul/fronthaul traffic could be estimated as described in a study performed by Next Generation Mobile Networks (NGMN) alliance\footnote{Here, related discussions have been deduced from a white paper by the NGMN Alliance titled as Small Cell Backhaul Requirements. Visit \url{https://www.ngmn.org/} for more information.}. As an example, an aggregated backhaul/fronthaul traffic $R_{agg}$ generated by $N$ SBSs can be estimated as $R_{agg}\;$=\;max$(N R_{busy}, R_{peak})$, where $R_{busy}$ denotes the average traffic during busy time for a single SBS when many users are being served by the respective cell and $R_{peak}$ denotes the peak cell throughput. In a largely populated areas, the aggregated backhaul/fronthaul traffic $R_{agg}$ is driven by data rate during busy times. In such a scenario, the FSO link can support up to $N=\lceil\frac{R}{R_{busy}}\rceil$ SBSs where $\lceil\cdot\rceil$ denotes the ceiling function.
\begin{table}[t!]
\caption{Summary of simulation parameters.}
\centering
\begin{tabular}{|@{}c@{}|@{}c@{}|}
\hline
Parameter & Value\\
\hline
Transmit power $(P_{t})$ & 200 mWatt\\
\hline
Pointing losses $(L_{poi})$ & 2 dB\\
\hline
Optical losses $(L_{opt})$ & 2 dB\\
\hline
Divergence angle $(\theta)$ & 1 mrad\\
\hline
Elevation angle $(\phi)$ & $45^\circ$ \\
\hline
Receiver radius $(r)$& 4 cm\\
\hline
Transmission wavelength $(\lambda) $& 1550 nm\\
\hline
NFP height ($h$) & 1 km - 20 km\\
\hline
Wind speed $(v)$& 21 m/s\\
\hline
Receiver sensitivity $(N_b)$ & 100 photons/bit\\
\hline
BER & $<10^{-9}$\\
\hline
Fog visibility $(V)$ & 50 m\\
\hline
Fog layer thickness $(\Delta d_{fog})$ & 50 m\\
\hline
Cloud attenuation $(L_{cloud})$& as proposed in \cite{Awan}\\
\hline
Rain rate $(R_{rain}) $& 50 mm/hour\\
\hline
Rain layer thickness $(\Delta d_{rain})$ & 1000 m\\
\hline
Planck's constant $(h_p)$ & $6.626\times 10^{-34}$ J-s \\
\hline
Speed of light $(c)$ & $3\times 10^{-8}$ m/s\\
\hline
\end{tabular}
\end{table}

\begin{NewBox}[t]
\textbf{Doppler Effect:} The relative motion between the NFPs and the SBSs may result in Doppler effect. Therefore, the vertical FSO transceiver should be equipped with a technique to compensate for Doppler effect. Some examples of these techniques, which have been proposed for space FSO, are Optical Phase-Lock loop (OPLL), Optical Injection Locking (OIL) technique, a combination of OPLL technique and OIL technique, and Optical Frequency Locked Loop (OPLL) \cite{Kaushal}. 
\end{NewBox}

The achievable data rate of a FSO link is given by \cite{Majumdar}
\begin{equation}
R=\frac{P_t\eta_t\eta_r 10^{\frac{-L_{poi}}{10}} 10^{\frac{-L_{atm}}{10}}A_R}{A_B E_p N_b}\:\:\:\:[\text{bits/s}],
\end{equation}
where $P_t$ denotes the transmit power, $\eta_t$ and $\eta_r$ stand for the optical efficiencies of the transmitter and receiver, respectively, $L_{poi}$ is the pointing loss measured in dB, $L_{atm}$ denotes the atmospheric attenuation 
due to rain, fog, cloud or turbulence measured in dB over the path length and given by $L_{atm}=L_{rain}+L_{fog}+L_{cloud}+L_{sci}$. 
Moreover, $E_p=h_pc/\lambda$ denotes the photon energy with $h_p$ denotes Planck's constant, $c$ denotes the speed of light and $\lambda$ stands for transmission wavelength. Finally, $N_b$ represents the receiver sensitivity in number of photons/bit. Unless otherwise specified, Table II summarizes the simulation parameters used in this article.
\begin{figure*}[t]
\begin{center}
\includegraphics[scale=.65]{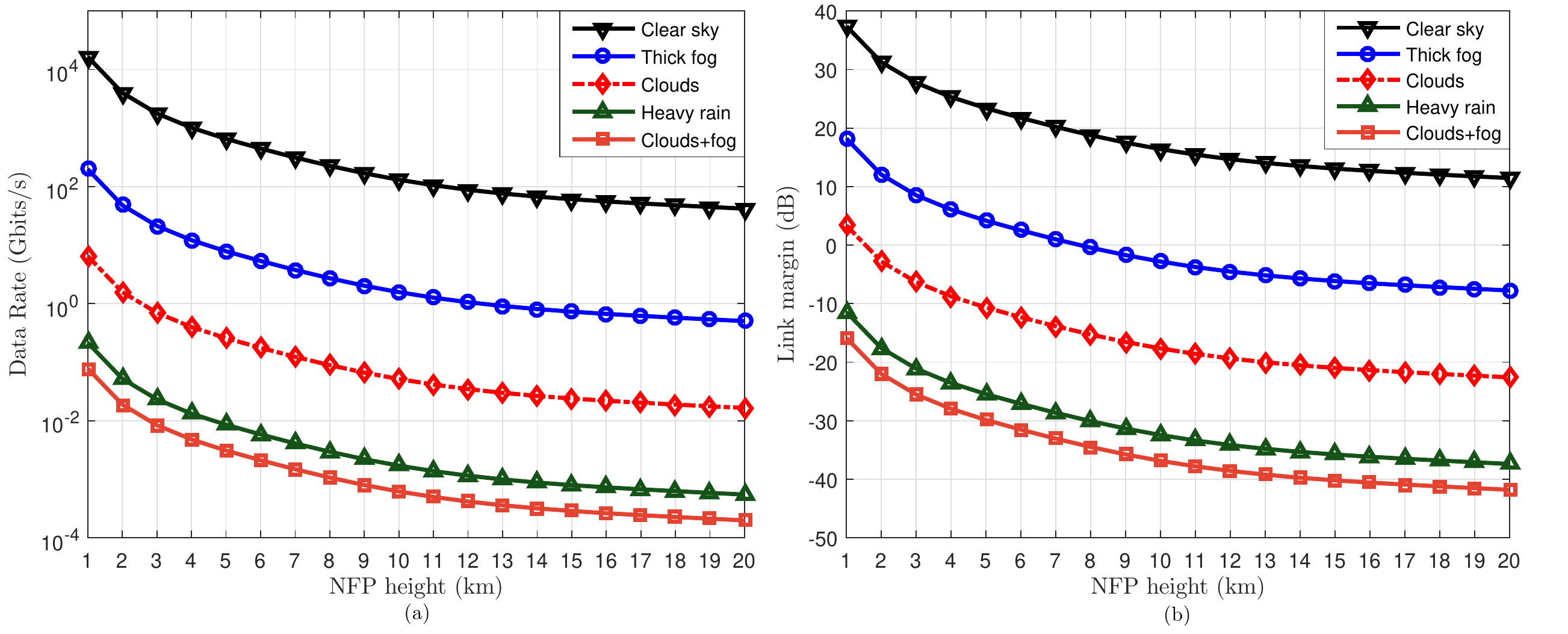}
\caption{\footnotesize Comparative performance summary of vertical FSO link for different weather conditions vs range of NFPs' altitude: (a) data rate, and (b) link margin.}
\label{fig2}
\end{center}
\end{figure*}

\begin{figure*}[t]
\begin{center}
\includegraphics[scale=0.65]{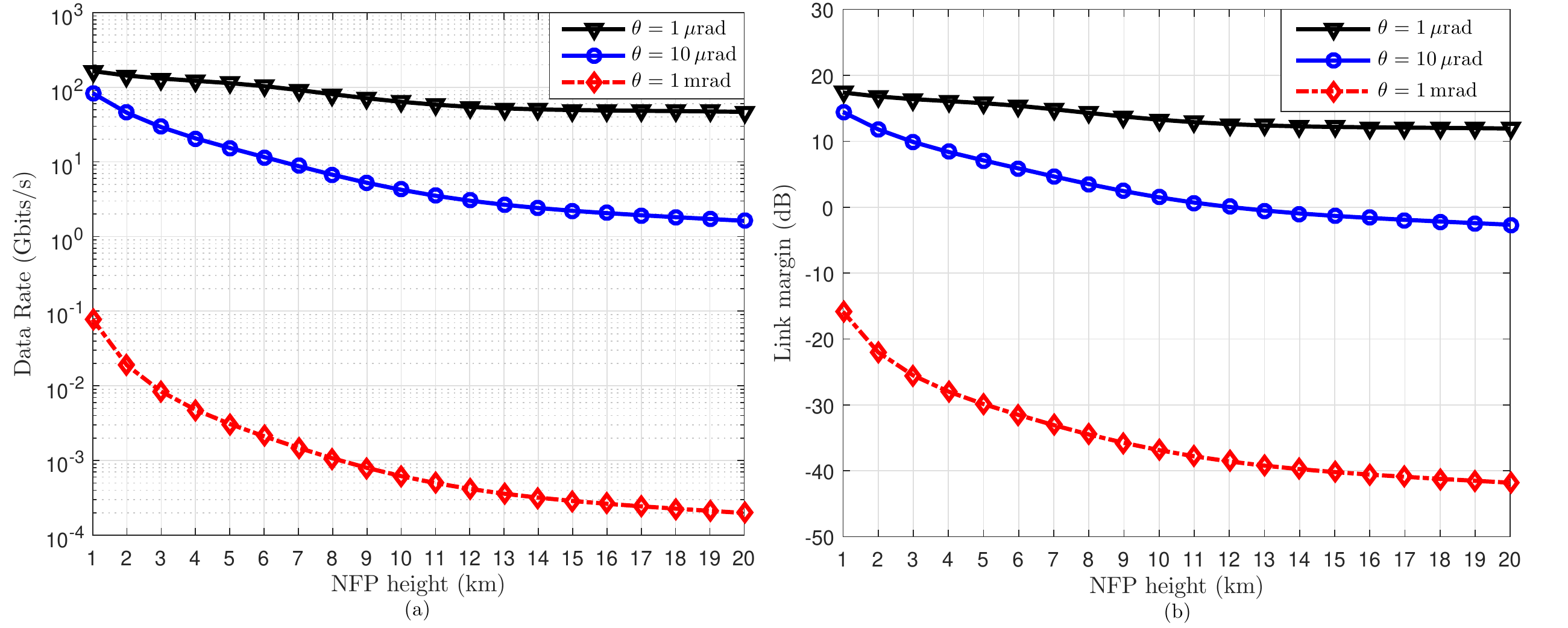}
\caption{\footnotesize Comparative performance summary of vertical FSO link for different divergence angles vs range of NFPs' altitude: (a) data rate, and (b) link margin.}
\label{fig3}
\end{center}
\end{figure*}

Fig. \ref{fig2} shows the achievable data rate and available link margin versus NFP altitude for different weather conditions. As shown in Fig. \ref{fig2}(a), with the increase in the altitude, the achievable data rate decreases for all weather conditions due to the increase in geometrical loss. It can also be seen that the vertical FSO link is mostly affected by the clouds because the wavelength of FSO beam is comparable to the size of the cloud particles (Mie scattering). Although fog also causes Mie scattering as clouds do, fog may not be a key issue for the vertical FSO links as for terrestrial  FSO links because the vertical FSO beam is vulnerable to fog for a relatively much shorter distance (fog layer thickness) while terrestrial FSO link is vulnerable to fog along the communication link. For all weather conditions except clear sky, the data rates at higher altitudes are relatively lower as compared to clear sky, e.g., at an altitude of 20 km, $R=$198 kbits/s, 0.5 Mbits/s and 42 Gbits/s for clouds and fog, heavy rain and clear sky, respectively. Fig. \ref{fig2}(b) illustrates the available link margin versus the NFP altitude for different weather conditions. As seen in this figure, 23 dB and 11.5 dB are available at an altitude of 5 km and 20 km, respectively, for a clear sky. Under cloudy, raining and foggy conditions, the FSO link fails because the received power is less than the sensitivity of the receiver, e.g., the available link margin is -42 dB for cloudy and foggy conditions at 20 km.

Fig. \ref{fig3} shows the achievable data rate and link margin for cloudy and foggy conditions (worst scenario) over the range of the NFP altitude for different divergence angles. As seen in Fig. \ref{fig3}(a), the achievable data rate can be improved by reducing the divergence angle, e.g., a data rate of  198 kbits/s, 1.6 Gbits/s, and 46 Gbits/s can be achieved with a divergence angle of 1 mrad, 10 $\mu$rad, and 1 $\mu$rad, respectively, at an altitude of 20 km. It can also be seen in Fig. \ref{fig3}(b) that reducing the divergence angle results in increasing the available link margin at the receiver, e.g., at an altitude of 20 km, a link margin of 12 dB is available at the receiver with a divergence angle of 1 $\mu$rad  compared to -2.6 dB (link failure) with a divergence angle of 10 $\mu$rad. 

\begin{figure*}[t]
\begin{center}
\includegraphics[ height=8.2cm, width=17cm]{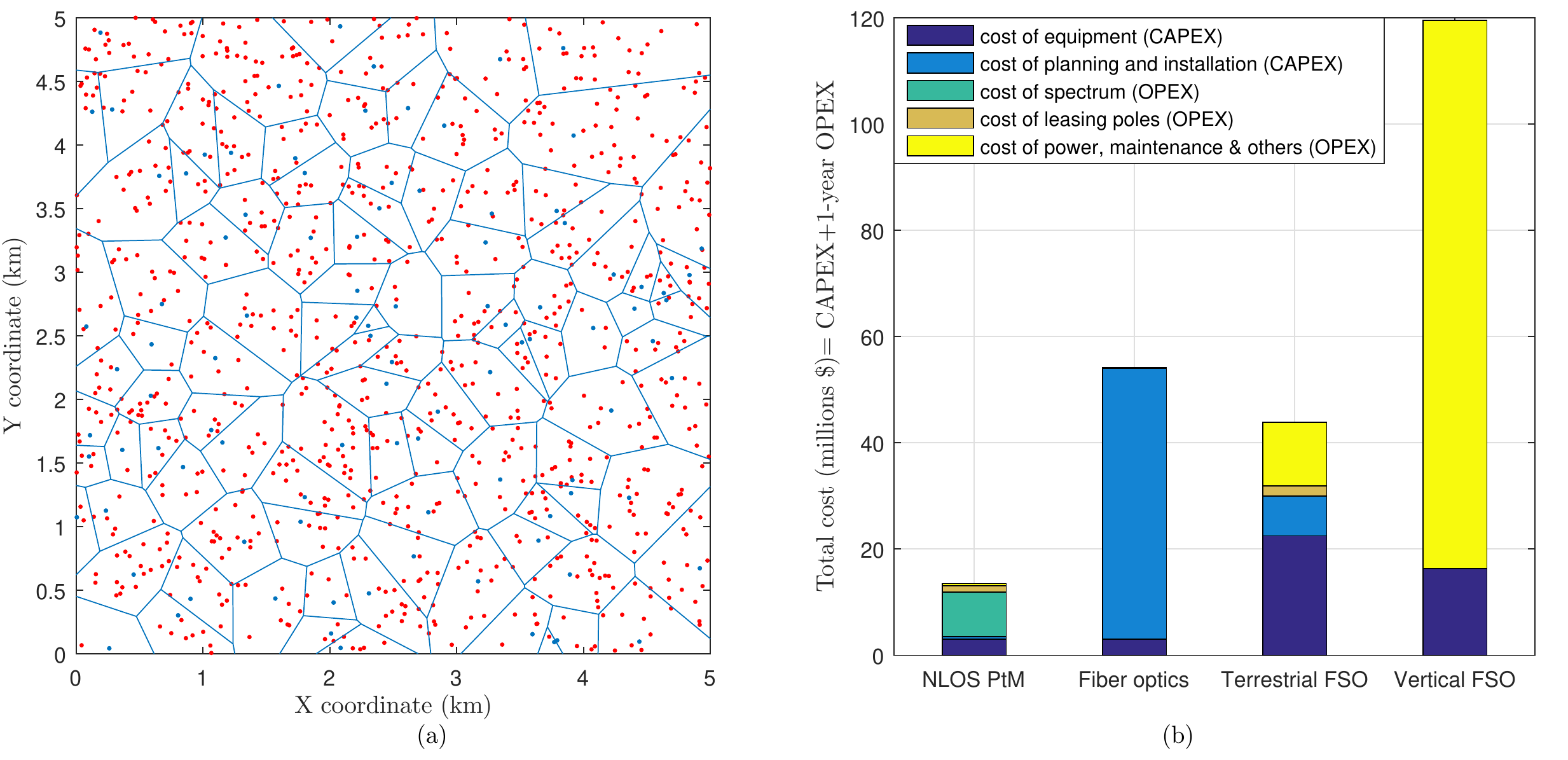}
\caption{\footnotesize Comparative summary of deployment of several backhaul/fronthaul technologies: (a) a snapshot of the typical Poisson distributed HetNet used to estimate the cost of backhaul/fronthaul links, and  (b) associated cost of several backhaul/fronthaul technologies in underlying HetNet.}
\label{fig4}
\end{center}
\end{figure*} 
\subsection{\textbf{Economics of Vertical System}}

The total cost of ownership (TCO) consists of the capital expenditures (CAPEX) and the operational expenditures (OPEX) of the backhaul/fronthaul network. While the CAPEX describes the cost of equipment, planning and installation, the OPEX describes the cost of spectrum, maintenance, power and fuel.

Assume a HetNet where 100 macro cells and 1000 small cells are deployed over an urban area of 5 km by 5 km. For RF NLOS PtM, fiber optics and FSO solutions, we assume that the small cells backhaul/fronthaul traffic is aggregated at an aggregation hub located at the macro-BS. For RF NLOS PtM, the configuration is assumed to be 1:4, i.e., the hub communicates with 4 remote backhaul modules located at the small cells. The cost of the hub and the remote backhaul module is assumed as \$4000 and \$2000, respectively \cite{Dahrouj}\footnote{The cost assumptions are based on the average prices in North America. However, they are applicable to the countries or regions of similar demographics.}. The installation cost for the hub and the remote backhaul module is assumed as \$270 and \$140, respectively. We also assume that the RF NLOS PtM solution operates in a 40 MHz licensed spectrum at the cost of \$0.007 per MHz per capita. The cost of leasing a pole is assumed as \$1250 per year and the power and maintenance cost is \$375 per year. For the optical fiber, the fiber optic cable is \$10 per meter and the installation cost is \$200 per meter while the power and maintenance cost is \$200 per link per year. For the terrestrial FSO, we assume that 50\% of the small cells have no LOS to the aggregation hubs, which is the case in urban areas, particularly, downtowns where the small cells are expected to be deployed near street levels rather than in clear spaces. Therefore, two or more terrestrial FSO links may be required to reach the small cell. The cost of the terrestrial FSO equipment is assumed as \$15,000 and the associated planning and installation cost is further assumed as \$5000. The cost of power and maintenance is \$8000 per link per year. For the proposed vertical solution, we assume 20 medium altitude long endurance (MALE) NFPs, which can carry heavy payloads for long endurance. The cost of the platform is assumed as \$50,000\footnote{Assumptions on costing of FSO equipment have been deduced from the discussions presented at the Panel US German Aerospace Round Table (UGART) on the next big thing in space, organized during an annual trade show and conference, Space Tech. Expo. Europe, Bremen, Germany, Nov. 2015.}. We also assume that the operating cost of each platform is \$859 per flight-hour \cite{GOA}.

Fig. \ref{fig4}(a) shows a snapshot of the typical Poisson distributed HetNet. Fig. \ref{fig4}(b) shows the cost for RF NLOS PtM, optical fiber, terrestrial FSO and FSO-based vertical systems for one year. It can be seen from the Fig. \ref{fig4}(b) that fiber optics has the highest deployment cost because of the high cost of digging and trenching in urban areas. On the other hand, the RF NLOS solution seems to be the most cost effective solution but it suffers from interference and low data rate due to spectrum sharing between the remote backhaul modules. Terrestrial FSO has a high deployment cost because LOS is not always available in urban areas. Therefore, more FSO equipment is required to reach the small cells. To reduce the deployment cost of terrestrial FSO solution, the aggregation hub can be implemented in the stratosphere. 
However, by considering the TCO, the vertical system has the highest cost at around \$120 million while the NLOS PtM, terrestrial FSO, and optical fiber solutions cost around \$14, \$44, \$55 million, respectively.

\section{Open Challenges and Future Research}

\subsection{\textbf{Implementation}} 

As previously discussed, the vertical FSO link is greatly affected by weather conditions. One possible approach is to design an adaptive algorithm that adjusts the transmit power according to weather conditions, e.g., under raining conditions, high power vertical FSO beams should be used while low power beams could be used under clear sky conditions. The algorithm may also adjust other system parameters such as divergence angle of the FSO transceiver to compensate the link degradation. The future implementation may also include a system optimization algorithm that can optimize the NFP placement, e.g., flying below clouds over negligible turbulence region. Another possibility is to use the Millimeter-wave spectrum (mm-wave) in a combination with FSO. Unlike FSO, mm-waves are not attenuated by fog. However, mm-waves are highly attenuated by water molecules such as rainfall \cite{Siddique}. To combat the advantages of both FSO and mm-waves, a hybrid FSO/mm-Wave can be considered. The system switches to FSO during rainy conditions when mm-wave transmission is not favorable due to the high rain attenuation and switches to mm-waves during foggy conditions. The proposed vertical backhauling/fronthauling system could also consider a hybrid FSO/RF as a potential alternative solution to overcome the link degradation under bad weather conditions \cite{Dahrouj}.

\subsection{\textbf{Cost}}
The proposed vertical backhaul/fronthaul network requires NFPs that can fly with heavy payloads (FSO systems) with long endurance. The UAVs that can fulfill these requirements are mainly designed for military missions. The operating cost of these UAVs is high, e.g., the Predator, which is a MALE UAV, costs around a thousand dollars per flight-hour \cite{GOA}. However, with the increasing interest of UAVs, either in military and civil applications, these UAVs are becoming less expensive which may reduce the cost of the vertical backhaul/fronthaul network. Another option is to use unmanned balloons as flying hubs. The balloons are often solar-powered and fly at a quasi-stationary position which could reduce the operational cost of the proposed system.

\subsection{\textbf{Safety and Regulatory}}

The regulation (safety, environment, etc.) to exploit the NFPs for such commercial use in future cellular networks is still underway. Several Canadian, US and European organizations have been working very closely to harmonize the regulatory approaches for flying platforms in their respective airspaces for commercial use including their usage for future cellular services. The regulatory processes include, but not limited to, issuance of flight operating licenses and air operator certificates to the cellular operators or any other commercial agency to authorize the flying of platforms and ensure that the platform flight is equipped with the adequate safety equipment for various environments and capable of managing the risk associated with the operation of the NFPs. The current focus of regulatory bodies is largely on the safety aspects of platform flight, particularly if they are to operate beyond LOS and in populated urban areas. 

\subsection{\textbf{NFP-Small Cell Association}}
With densely deployment of small cells, the NFP-small cell association becomes more challenging in underlying vertical systems. Each NFP has a constraint on the payload it can carry. Therefore, each NFP can serve at most a particular number of small cells because of the limited number of FSO transceivers it can fly with. Furthermore, the NFP-backhaul link that forwards the traffic from the NFP node to the mother NFP or vice versa could be strictly constrained by the limited capacity of that link. Therefore, even if the NFPs carry enough FSO transceivers, the number of small cells served by each NFP could be limited by the capacity of the NFP-backhaul link. It is clear that successful integration of the proposed vertical fronthaul/backhaul system relies on advanced optimization of sophisticated design parameters such as the optimal number of NFPs required to provide fronthaul/backhaul to the small cells and their association with the small cells considering the typical factors, e.g., payload, achieving data rate and backhaul data rate between the NFP and ground station.

\subsection{\textbf{Security and Privacy}} 
Security and privacy for NFPs are considered as fundamental requirements for the proposed vertical system. In general, the use of NFPs for the future cellular networks could be limited due to following two reasons: 
\begin{itemize} 
\item The security risk that the NFPs could be hijacked/sabotaged may result in disruption or complete failure of cellular systems and services. Operators, cellular vendors or any other commercial agency intending to exploit NFPs for commercial services, are required to integrate advanced control mechanisms to ensure the security of the NFPs and their operation. 
\item The privacy and data protection of the cellular network entities that are connected to the NFPs could also be at risk due to inadequate security measures. Some regulatory measures could be taken to maintain additional privacy and protection of the cellular entities in the NFP-based network such as prohibiting the NFPs from flying over critical and unauthorized infrastructure and from carrying payload that could potentially collect personal data and information.
\end{itemize}

\section{Conclusion}
In this article, we investigated a vertical  framework to backhaul/fronthaul SBSs via NFPs such that these NFPs relay the SBSs to the core network. The proposed system is envisioned to be deployed as a complementary solution to the terrestrial solutions to offer highly reliable backhaul/fronthaul system for challenging ultra-dense SBSs deployments. The efficacy of the proposed system has been investigated in terms of link budget and achievable data rate for a single FSO link under different weather conditions.

Simulations have shown that the key challenge is the high path loss under some weather conditions, in particular, clouds accompanied by fog. However, the performance can be improved significantly and rates in the order of multi Gbits/s can be achieved by reducing the divergence angle. The economics of the system has also shown that the vertical system has a high TCO compared to terrestrial backhaul/fronthaul networks. However, the vertical backhaul/fronthaul system and the associated equipment is a mature technology, and their cost is expected to decrease faster than the existing terrestrial solutions. 


\section*{Acknowledgements}
This work was supported in part by the Ministry of Higher Education  and Scientific Research (MOHESR), Libya,
through the Libyan-North American Scholarship Program, in part by  Huawei Technologies Canada, and in part by the Ontario
Ministry of Economic Development and Innovations Ontario Research Fund  — Research Excellence Program.

\balance
{
\bibliography{IEEEfull,RefList}
\bibliographystyle{IEEEtran}}

\end{document}